\documentclass[prod]{jfmfr}
\usepackage{graphicx}
\usepackage{epstopdf, epsfig}
\usepackage{bm}
\usepackage{hyperref}
\usepackage{siunitx}
\usepackage{layouts}
\usepackage{amsmath}  
\usepackage[dvipsnames]{xcolor}
\usepackage{ulem}

\shorttitle{Phoresis in cellular flows}
\shortauthor{R. Volk, M. Bourgoin, C.-E. Br\'ehier and F. Raynal}

\title{Phoresis in cellular flows:\\ 
from enhanced dispersion to blockage}

\author{Romain Volk\aff{1} 
   \corresp{\email{romain.volk@ens-lyon.fr}},
     Micha\"el Bourgoin\aff{1},
  Charles-\'Edouard Br\'ehier\aff{2}\\
 \and Florence Raynal\aff{3}
   \corresp{\email{florence.raynal@ec-lyon.fr}}
   }

\affiliation{\aff{1}Laboratoire de Physique, ENS de Lyon, CNRS, 69364 Lyon CEDEX 07, France
\aff{2}Institut Camille Jordan, Univ Lyon, Universit\'e Claude Bernard Lyon 1, CNRS, F-69622 Villeurbanne CEDEX, France
 \aff{3}Laboratoire de M\'ecanique des Fluides et d'Acoustique, Univ Lyon, \'Ecole Centrale Lyon, INSA Lyon, Universit\'e Claude Bernard Lyon 1, CNRS, F-69134 \'Ecully, France. }

\begin{document}

\date{Received 25 March 2022; revised 12 July 2022; accepted 20 August 2022} 
\pubyear{2022} 
\volume{948, A42}
\pagerange{A42-1--A42-19}
\doi{jfm.2022.730}

\maketitle

\begin{abstract}

In this article, we study numerically the dispersion of colloids in a two-dimensional cellular flow in the presence of an imposed mean salt gradient. 
Owing to the additional scalar, the colloids do not follow exactly the Eulerian flow field, but have a (small) extra-velocity proportional to the salt gradient, $\mathbf{v}_\mathrm{dp}=\alpha\nabla S$, where $\alpha$ is the phoretic constant and $S$ the salt concentration. We study the demixing of an homogenous distribution of colloids and how their long-term mean velocity $\mathbf{V_m}$ and effective diffusivity $D_\mathrm{eff}$ are influenced by the phoretic drift. 
We observe two regimes of colloids dynamics depending on a blockage criterion $R=\alpha G L/\sqrt{4 D_cD_s}$, where $G$ is the mean salt gradient amplitude, $L$ the length scale of the flow and $D_c$ and $D_s$ the molecular diffusivities of colloids and salt. 
When $R<1$, the mean velocity is strongly enhanced with $V_m \propto \alpha G \sqrt{\Pen_s}$, $\Pen_s$ being the salt P\'eclet number. 
When $R> 1$, the compressibility effect due to the phoretic drift is so strong that a depletion of colloids occurs along the separatrices inhibiting cell-to-cell transport.

\end{abstract}



\section{Introduction}


The transport of a passive contaminant or a magnetic field in flows with closed streamlines has been the focus of many studies in the past few decades, see, for instance, \citet{solomon1988passive,bib:shraiman1987,soward_1987,young1989anomalous}. 
In such configurations, it has been shown that the transport of Brownian particles is greatly enhanced by advection: their dynamics at long time becomes diffusive, with the ratio of their effective diffusivity to their molecular diffusivity growing as the square root of the P\'eclet number, $\Pen=UL/D$, where $U$ is the velocity scale, $L$ its the length scale, and $D$ the molecular diffusivity of the particles. 
Since these pioneering studies, modified versions of this problem have been addressed in the context of non-tracer particles {\it i.e.}, particles which do not strictly follow the fluid motions because of their response time to flow modifications \citep{bib:maxey1987}. 
The question of long time dynamics of very heavy, weakly inertial, particles settling under the action of gravity was addressed by \cite{PAVLIOTIS2005161} and \cite{Afonso_2008} who derived expressions for the mean particle settling velocity, which was found to be larger than the settling velocity in a quiescent fluid. 
More recently, such study was extended to the case of arbitrary density ratio by \cite{renaud_vanneste_2020}; 
in particular, the authors pointed out the role of compressibility effects due to particle inertia, which can lead to reduced transport in the case of light particles that tend to be trapped in vortical flow regions \citep{maxey1987gravitational,wang1993settling}. 
Reduction of effective diffusivity was also observed numerically by \cite{bib:li_etal2021} who studied the case of passive Brownian particles submitted to the action of a mean transverse force in an array of vortex similar that of \cite{solomon1988passive}, and in the case of active particles moving in a laminar vortical flow as transport barriers develop \citep{bib:berman_etal2021}.

Phoretic particles such as colloids are another class of particles which do not follow exactly the fluid motion;  indeed, they are submitted to a drift velocity that originates from  heterogeneities in the background of chemical (diffusiophoresis) or thermal (thermophoresis) origin. 
\cite{bib:Anderson1989} showed that the drift velocity for diffusiophoresis is of the type $\mathbf{v}_\mathrm{dp} = D_\mathbf{dp} \nabla \log S$, where $D_\mathrm{dp}$ is the diffusiophoretic mobility and $S$ is the concentration in background chemical (\textit{e.g.} salt), whereas
\cite{bib:gupta_etal2020} determined experimentally that the dependence can be rather more complex (in particular, $D_{dp}$ may depend on the salt concentration). 
These observations are in accordance with a recent more complex expression for the phoretic velocity proposed in \cite{bib:menolascina_etal2017,bib:salek_etal2019}. 
Thus, depending on the species considered and its concentration, a drift velocity of the type $\mathbf{v}_\mathrm{dp}=\alpha\nabla S$ can also be observed in practice (corresponding to $D_\mathrm{dp}\propto S$), as discussed in \cite{bib:chu_etal2022}; 
this is the expression we consider in this article. 
Note finally that in a view of simplicity, in the following we mainly refer to diffusiophoresis. 
However, other mechanisms like chemotaxis (for instance, movement of a motile organism in a direction corresponding to a gradient concentration of a nutriment) lead to a similar expression for the drift velocity, but with a much larger drift coefficient $\alpha$, as also discussed in \citet{bib:chu_etal2022}.

Mixing of phoretic particles in the presence of salt heterogeneities was first studied experimentally in a $\Psi$-shaped channel \citep{bib:Abecassisetal2009}, and later in chaotic advection at the micro- or macro-scale \citep{bib:Deseigneetal2014,bib:Maugeretal2016}. 
In particular, the authors showed that mixing is delayed when the colloids are introduced together with the salt, and enhanced when introduced in salted water. 
In this context, it has been shown that although the drift velocity is usually very weak as compared with the flow velocity, mixing is strongly modified because, due to diffusiophoresis, the particle velocity field is compressible
\citep{bib:Volketal2014,bib:raynaletal2018,bib:raynal_volk2019,bib:chu_etal2020,bib:chu_etal2021}. 
One may thus wonder how the long time transport of such phoretic particles by a cellular flow is modified when a mean scalar gradient such as salt or temperature is imposed to the system. 
It can be anticipated that in the large time regime, a stationary state is reached and we are interested here in the behaviour of the effective velocity and diffusivity in this asymptotic regime: for instance, we wonder whether the average colloids velocity will be enhanced or reduced as compared with the case without advection. 
Answering this question is not trivial because the presence of strongly localised gradients may lead to enhanced cell-to-cell transport as well as arrested transport as observed previously in a simpler geometry \citep{bib:chu_etal2020}.

In this article, we study the dispersion of colloids in a 2-dimensional cellular flow in the presence of an additional scalar (which we call salt) forced with an imposed mean salt gradient. The flow satisfies
\begin{equation}
\mathbf{u}(x,y)= u_0 (\sin kx \cos ky, -\cos k x \sin ky)\,,
\label{eq:cellular_velocity_field}
\end{equation}
with $k=\pi/L$. 
It is $2L$ periodic in space and has closed streamlines in square cells of side $L$. 
It is a good model for Rayleigh-B\'enard flows close to the instability threshold \citep{solomon1988passive}; the same flow was also realised using an electrolyte, a network of magnets and a DC current to drive the flow using the Lorentz force in order to address particle settling \citep{bib:bergougnoux_etal_2014}. 

Owing to the presence of salt gradients, the colloids do not follow exactly the Eulerian flow field $\mathbf{u}$, but have a (small) extra velocity of the type
\begin{equation}
\mathbf{v}_\mathrm{dp}=\alpha \nabla S\,,
\label{eq:vdp_def}
\end{equation}
where $S$ is the salt concentration and $\alpha$ is the (small) phoretic coefficient taken constant and independent of $S$.
Salt and colloids concentration fields ($S$ and $C$) satisfy the advection diffusion equations 
\begin{eqnarray}
    &\partial_t S + \nabla \cdot  S \mathbf{u} = D_s\nabla^2 S \,, \label{eq:salt}\\
    &\partial_t C + \nabla \cdot C (\mathbf{u}+ \mathbf{v}_{\mathrm{dp}}) = D_c\, \nabla^2 C\,,\label{eq:colloids}
\end{eqnarray}
where $D_s$ and $D_c$ are the molecular diffusion coefficients of salt and colloids, respectively. 
Note the one-way coupling: from equation \eqref{eq:salt} the salt evolves freely, whilst colloids are coupled to salt via the phoretic velocity $\mathbf{v}_\mathrm{dp}$ (equations \eqref{eq:colloids} and \eqref{eq:vdp_def}).
Note also that equation \eqref{eq:colloids} is linear in $C$; we suppose that the mean colloids concentration $C_0$ is such that the mean distance between colloids is sufficiently large so that colloid--colloid interactions can be neglected.

The salt concentration is forced by an imposed mean salt gradient in the $x$-direction
\begin{equation}
   \mathbf{G}=\langle \nabla S\rangle=G\, \mathbf{e}_x\,,
   \label{eq:imposed_G}
\end{equation}
where $\langle \cdot\rangle$ represents a spatial average over a square of side $2L = 2 \pi/k$. 
Using the $\pi$-theorem \citep{bib:vaschy,bib:buckingham} shows that the problem is governed by three independent non-dimensional parameters: 
we choose the salt and colloids P\'eclet numbers $\Pen_s=u_0 L/D_s$ 
and $\Pen_c=u_0 L/D_c$ respectively, and a new additional  number, a criterion of blockage, which will be shown to be the ratio $R=\alpha GL/\sqrt{4D_cD_s}$, see section \ref{sec:blockage}. 
Therefore, without loss of generality we can fix $u_0$, $L$ and $G$ and only vary $\alpha$, $D_s$ and $D_c$. 
In the following, we set $u_0=1$, $L=\pi$ and $G=1$ for the numerical simulations. 

By means of high-resolution numerical simulations performed both in the Eulerian and Lagrangian frameworks, we study the demixing of an homogenous distribution of colloids and how their mean velocity $\mathbf{V}_m$ and effective diffusivity $D_\mathrm{eff}$ (obtained in the long time regime, see section \ref{sec:colloids}) are influenced by the phoretic drift for a wide range of parameters. 

This paper is organised as follows. 
In Section \ref{sec:salt} we address the well-documented case of mixing of diffusing salt in a cellular flow in the presence of an imposed mean salt gradient $\mathbf{G}$, which converges towards a stationary state. 
In Section \ref{sec:colloids} we describe the demixing of diffusing colloids in this stationary salt concentration field; 
here again the colloids concentration field reaches a stationary state. 
We consider Eulerian simulations, and explain the typical concentration fields observed. 
Then we consider Lagrangian simulations for higher colloids P\'eclet number $\Pen_c$, and study the net mean velocity $V_m$ reached by the colloids, together with their effective diffusion; 
in particular, we show that, depending on the parameters considered, $V_m$ can be much higher than what could be expected, but also much less, up to vanishing, a situation referred to as ``blockage''. 
Finally in Section \ref{sec:enhanced_disp_blockage} we go into  more detail concerning the mean colloids velocity analysis. 
We deduce the order of magnitude of $V_m$ when no blockage occurs, obtain the condition of blockage and compare with numerical simulations in a wide range of parameters. 


\section{Case of salt}
\label{sec:salt}
The transport of a scalar or a vector in a cellular flow has been the focus of many studies in the past decades \citep{bib:shraiman1987,soward_1987,young1989anomalous,Afonso_2008,renaud_vanneste_2020}. 
We propose here to briefly outline the subject in the context of an imposed mean gradient which corresponds to the original configuration of \cite{bib:shraiman1987}.

The salt concentration $S(x,y,t)$ evolves independently of the colloids and is forced by the imposed gradient $\mathbf{G}=(G_x,G_y)$ following equations \eqref{eq:salt} and \eqref{eq:imposed_G}. 
In practice, we solve numerically the equation for $S'=S-\mathbf{G} \cdot \mathbf{x}$, which satisfies the equation
\begin{equation}
    \partial_t S' + \nabla \cdot  S' \mathbf{u} = D_s\nabla^2 S' - \mathbf{G} \cdot \mathbf{u} \,.
    \label{eq:s2}
\end{equation}
Because the velocity field is periodic in space with zero average 
\begin{equation}
    \langle \mathbf{u} \rangle = \frac{1}{4 L^2} \iint_{[0,2L]^2} \mathbf{u}(x,y)\, dxdy = 0\,, 
\end{equation}
equation \eqref{eq:s2} for $S'$ can be solved with periodic boundary conditions with $\langle S' \rangle = 0$ \citep{bib:Holzer1994}. This ensures that the total salt concentration field $S=S'+\mathbf{G} \cdot \mathbf{x}$ satisfies boundary conditions compatible with the imposed mean gradient:
\begin{eqnarray}
 S(x+2L,y)&=&S(x,y)+2G_x L\\
 S(x,y+2L)&=&S(x,y)+2G_y L\,.
\end{eqnarray}
In the following, without any loss of generality, we set $L=\pi$ for the numerical simulations.
The equation for $S'$ is solved in spectral space in a square domain $(x,y) \in [0,2 \pi]^2$ using the method described in \cite{bib:Volketal2014} with a mean gradient in the $x$ direction ($\mathbf{G} = (G,0)$). 
Because equation \eqref{eq:s2} is linear, $S'$ is proportional to $G$ so that we set $G=1$ without loss of generality. 
We set $u_0=1$ and adjust the spatial resolution to $256^2$ for P\'eclet number $\Pen_s=u_0 L/D_s \in [30,300]$ and $512^2$ for $\Pen_s=u_0 L/D_s \in [300,3000]$. 
Such very high resolutions are chosen to ensure high precision when interpolating the salt gradient in real space which is needed for the Lagrangian dynamics of colloids. 

Integrating in time from an initial condition $S'=0$, the salt concentration develops strongly localised gradients and reaches a stationary state ($\partial_t S=0$) thanks to the stationarity of the forcing term $-G\,u_x$ (equation \ref{eq:s2}). Such localisation can be understood in the limit of infinite P\'eclet numbers.
Indeed, in the case of a non-diffusing scalar ($D_s=0)$, equation \ref{eq:salt} leads to $\mathbf{u}\cdot\nabla S=0$, which admits a solution for which $\nabla S$ is zero almost everywhere. Given the forcing, it implies that $S$ is piecewise constant and increases by jumps by a value $GL$ in the direction $x$ of the forcing; the jump occurs at the vertical boundary of a cell, where the velocity is vertical and perpendicular to the salt gradient so that $\mathbf{u}\cdot\nabla S=0$. 

\begin{figure}
    \centering
    \includegraphics{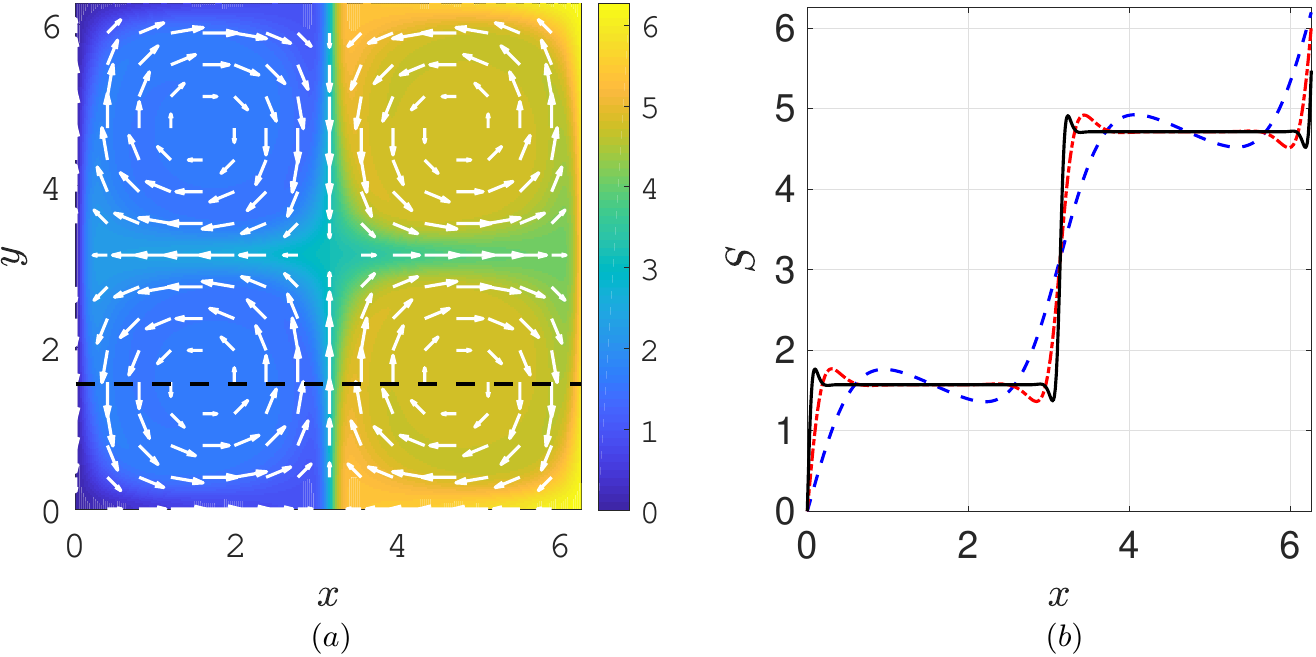}
    \caption{Salt case, with a positive mean imposed gradient $\mathbf{G}=\pi/L\,\mathbf{e}_x$ ($L=\pi$). $(a)$: Two-dimensional concentration field for $\Pen_s=314$; the concentration is almost constant in a given cell. $(b)$: One-dimensional cut along the dotted line in figure \ref{fig:salt_eulerian}\,$a$ $(y=\pi/2)$ for three different salt P\'eclet numbers: -~-~-: $\Pen_s=31.4$; -$\cdot$-$\cdot$-, $\Pen_s=314$; \textbf{---}, $\Pen_s=3\,141$; for $\Pen_s\ge 314$, the profile is flat except for a very small layer around the vertical separatrix where the concentration jumps abruptly.}
    \label{fig:salt_eulerian}
\end{figure}
This is indeed the picture that we obtain at large P\'eclet number, as shown in figure \ref{fig:salt_eulerian}, which displays the Eulerian salt concentration field obtained for $\Pen_s=314$ (Fig \ref{fig:salt_eulerian}\,$a$) together with a one-dimensional horizontal profile at mid-height of a cell for different P\'eclet numbers (Fig \ref{fig:salt_eulerian}\,$b$): 
the concentration is almost constant in a given cell, and jumps abruptly (but on a finite slope) by a value $\Delta S=GL$ from one cell to the next one in the $x$-direction. 

The typical width $\ell_s$ on which the concentration increases by a value $\Delta S$ (the salt Batchelor scale \citep{bib:Batchelor59}) results from a competition between contraction and diffusion through the equation of salt gradient $\vec{\gamma}^s=\nabla S$:
\begin{equation}
\frac{D\gamma_i^s}{Dt}= -\partial_i u_j\,\gamma^s_j +D_s\nabla^2 \gamma^s_i\,,
\label{equation_gradient_s}
\end{equation}
where $D/Dt=\partial_t+u_k\partial_k$ is the material derivative. 
This horizontal gradient is built by bringing the scalar to the hyperbolic point $(\pi,0)$ by contraction by the velocity field along the horizontal separatrix ($i=j=x)$; because the concentration field is stationary its width is set when the two terms on the right-hand side of equation (\ref{equation_gradient_s}) are of the same order of magnitude \citep{bib:raynalgence1997}, that is, $u_0/L\sim D_s/\ell_s^2$, therefore
\begin{equation}
\ell_s\sim \frac{L}{\sqrt{\Pen_s}}\,.
\label{eq:ell_s}
\end{equation}
Indeed, as the horizontal velocity reverses near the point $(\pi,0)$, the advection term $\mathbf{u}\cdot \nabla \gamma_x$ vanishes when averaged over the width $\ell_s$ while the contraction term remains.
This concentration gradient is then advected upwards by the velocity field along the vertical separatrix $x=L=\pi$. 
It reaches the top of the cell after a time $\tau^s_\mathit{adv}\sim L/u_0$, time after which it is dissipated by diffusion because the characteristic dissipation time is $\tau^s_\mathit{diff}\sim \ell_s^2/D_s$, and $\tau^s_\mathit{adv}=\tau^s_\mathit{diff}$ because of equation  \eqref{eq:ell_s}. 
Thus, this concentration discontinuity is only visible around the vertical separatrix, with an associated gradient
\begin{equation}
\vec\gamma^s\sim \frac{\Delta S}{\ell_s}\,\mathbf{e}_x\sim G\,\sqrt{\Pen_s}\,\mathbf{e}_x\,.
\label{eq:salt_grad_separatrix}
\end{equation} 
This scenario supposes that the salt can be transported on a large scale before diffusing, \textit{i.e.} $\Pen_s\gg 1$, in agreement with figure \ref{fig:salt_eulerian}\,$b$. 


\section{Case of colloids}
\label{sec:colloids}
In the following we suppose that the salt has reached its stationary state, and focus on the dynamics of the colloids which are advected by the velocity field $\mathbf{v}=\mathbf{u} + \mathbf{v}_{\mathrm{dp}}$ (with $\mathbf{v}_\mathrm{dp}=\alpha \nabla S$) and diffuse with a diffusion coefficient $D_c$. 
Starting from a \textit{uniform distribution}, colloids will demix because $\nabla \cdot \mathbf{v}_\mathrm{dp} \neq 0$ \citep{bib:Volketal2014} and reach a non-uniform stationary state that we will characterise thereafter.

This problem can be addressed either in the Eulerian or in the Lagrangian frameworks. 
On the one hand, one may use the Lagrangian approach by solving the stochastic differential equation governing the position $\mathbf{X}_c$ of the colloids:
\begin{equation}
    d\mathbf{X}_c = \bigl(\mathbf{u}(\mathbf{X}_c)+ \alpha \nabla S(\mathbf{X}_c)\bigr) dt + \sqrt{2 D_c}\, d\mathbf{W}\,,
    \label{eq:lag}
\end{equation}
where $\mathbf{W}$ is a two-dimensional Wiener process \citep{VANKAMPEN2007}. 
Such framework is used in the following in cases when the colloid P\'eclet number, $\Pen_c = u_0L/D_c$, is very large, in order to compute the mean velocity of the colloids and their effective diffusivity at long times $t \in [\tau_c/10, \tau_c]$, where $\tau_c=L^2/D_c$ is the diffusion time of the colloids. 
In practice, we compute a three cubic periodic interpolant of the salt gradient, resolved with very high precision, for each value of $Pe_s$, which is then used for all cases when varying $\alpha \in [0,~ 10^{-2}]$ and $D_c \in [10^{-6},  ~10^{-1}]$; starting from at least $\mathcal{O}(10^4)$ uniformly distributed colloidal particles, we solve equation \eqref{eq:lag} numerically for each trajectory by splitting the deterministic and stochastic parts at each time step. 
We use a fourth-order Runge-Kutta method for the deterministic part with $dt=10^{-2}$ and a standard Euler scheme Euler scheme for the stochastic part \citep{bib:higham525}. 


On the other hand, the problem may also be addressed in an Eulerian framework by investigating the colloids concentration field $C$, as $C(\mathbf{x},t)/\langle C\rangle$ is the probability of finding a colloid at location $\mathbf{x}$ at time $t$. 
The concentration $C$ is coupled with that of the salt through the diffusiophoretic velocity $\mathbf{v}_\mathrm{dp}=\alpha \nabla S$. 
We use the Eulerian approach and solve the partial differential equation \eqref{eq:colloids} in the case of moderate P\'eclet numbers ($Pe_c \leq 3\cdot 10^{4}$, $D_c \in [10^{-4},~ 10^{-1}]$) using the same code as  for the salt; without loss of generality we set the initial uniform condition $C(x,y,t)=C_0=1$ as the equation is linear in $C$. 
We run the simulation until $C$ has reached the stationary regime ($\partial_t C=0$).

In practice in this study, we restricted Eulerian simulations to cases with moderate colloids P\'eclet numbers: 
indeed, when the P\'eclet number increases, the colloid concentration fronts become steeper, even in the absence of diffusiophoresis. 
This is also the case when blockage occurs. 
Because steep fronts can only be caught at a very high numerical price with a spectral code, we limited the Eulerian cases to $\Pen_c\le 3100$. 
We tested the equivalence between the two methods on a range of intermediate P\'eclet numbers by comparing the mean Lagrangian colloid velocity to its Eulerian counterpart $\langle C (\mathbf{u}+ \alpha \nabla S)\rangle$: 
we successfully obtained the same value with both methods for all cases tested. 

When dealing with diffusiophoresis, it is usual to consider separately the \textit{salt-attracting} case where the colloids move toward the salt gradient ($\alpha>0$), and reversely the \textit{salt-repelling} case ($\alpha<0$) \citep{bib:Deseigneetal2014,bib:Volketal2014}. 
Indeed, changing the sign of the diffusiophoretic parameter may fully change the physics and the scaling involved \citep{bib:raynal_volk2019,bib:chu_etal2020}. 
This is not the case here: 
the stationary colloids concentration field is actually invariant under transformation $\alpha\rightarrow -\alpha$, $x\rightarrow -x$ and $y\rightarrow -y$, see Appendix \ref{app:alpha}.
This implies that the physics of the problem is unchanged when considering salt-attracted or salt-repelled colloids. 
Thus without loss of generality we set 
\begin{equation}
    \alpha\ge 0\,.
\end{equation}
Finally, as the drift velocity ($\mathbf{v}_\mathrm{dp} = \alpha \nabla S$) is linear in $\alpha\,G$, equations are unchanged if the imposed salt gradient $\mathbf{G}$ is reversed and $\alpha$ changed into $-\alpha$. 

Because colloids are expected to have a small diffusion coefficient as compared with the salt, we consider only cases for which 
\begin{equation}
    D_c \ll D_s
\end{equation}
(equivalently, $\Pen_c \gg \Pen_s$).
In the case $\alpha>0$, the colloids start to demix due to diffusiophoresis, and gradients of colloid concentration appear. 
In the stationary state, the width $\ell_c$ of these gradients is governed by the competition between the dominant mechanism of gradient creation (contraction either by the velocity field or diffusiophoresis) and destruction by diffusion. 
In the case when contraction by the velocity field is dominant, we obtain as for the salt a heterogeneity of width $\ell_c=L/\sqrt{ \Pen_c}$ along the vertical separatrix, with  $\ell_c\ll\ell_s$. 
When diffusiophoresis is dominant, it acts as an additional mechanism of creation of gradient, which leads to an even smaller Batchelor scale $\ell_c$. 
Finally, we have whatever $\alpha$
\begin{equation}
\ell_c\ll\ell_s\,.
\label{eq:ell_c_ll_ell_s}
\end{equation}

\subsection{Eulerian results: colloidal concentration field}

The salt field is uniform, except in the vicinity of the vertical separatrices where there is a very strong salt gradient. 
It is therefore understood that far from the separatrices, diffusiophoresis does not operate and the colloids move only along the current lines;
as a result, the colloids concentration remains constant in the central vortex.  
Because non-trivial behaviour takes place around the vertical separatrices, we consider the equation of evolution of the gradient in colloids $\vec\gamma^c=\nabla C$ \citep{bib:raynal_volk2019}
\footnote{Compared with our previous work \citep{bib:raynal_volk2019}, the diffusiophoretic velocity is here proportional to the salt gradient.}:
\begin{equation}
\frac{D\gamma^c_i}{Dt}
= \underbrace{-\partial_i u_j\,\gamma^c_j}_{(a)} +\underbrace{D_c\nabla^2 \gamma^c_i}_{(b)}
-\underbrace{\alpha \partial_j \gamma^c_i\partial_j S}_{(c)}
-\underbrace{\alpha \gamma^c_j\partial_i\partial_j S}_{(d)}
-\underbrace{\alpha  \gamma_i^c \partial_j^2 S}_{(e)}
-\underbrace{\alpha C\partial_i\partial_j^2 S}_{(f)}\,.
\label{equation_gradient_c}
\end{equation}
Let us evaluate each of those terms on the separatrix: 
\begin{eqnarray}
(a) &\sim& u_0\gamma^c /L\\
(b) &\sim& D_c \gamma^c/\ell_c^2\\
(c) &\sim&  \alpha  \gamma^c\Delta S/(\ell_c \ell_s)\\
(d) &\sim& \alpha  \gamma^c\Delta S/\ell_s^2 \ll (c)\\
(e) &\sim& \alpha  \gamma^c\Delta S/\ell_s^2 \ll (c)\\
(f) &\sim& \alpha  \gamma^c \Delta S\ell_c/\ell_s^3\ll (c)\,,
\end{eqnarray}
where the three last terms are negligible compared with term $(c)$ because of relation \eqref{eq:ell_c_ll_ell_s}.
As the velocity field $\mathbf{u}$ is perpendicular to the salt or colloids gradients along the vertical separatrix, as in Section \ref{sec:salt} the left hand-side term of equation (\ref{equation_gradient_c}) is zero in this stationary regime.
We consider the case when the mechanism of creation of gradients by diffusiophoresis is stronger than by advection, which gives $ (c)\gtrsim (a) $.
The width of the heterogeneity in colloids is then given by the competition between the creation of gradients by diffusiophoresis, and its destruction by diffusion ($(c)\sim(b)$).  
Therefore, we obtain
\begin{equation}
\ell_c\sim\frac{D_c\ell_s}{\alpha\Delta S}\sim \frac{D_c}{\alpha G \sqrt{\Pen_s}}  \,,
\label{eq:ell_c}
\end{equation}
given equation \eqref{eq:salt_grad_separatrix}. 
Combining condition $(c)\gtrsim(a)$ with scaling \eqref{eq:ell_c} gives the following condition for the parameter $\alpha$ to ensure diffusiophoresis has a stronger effect than advection:
\begin{equation}
\alpha\gtrsim\frac{\sqrt{D_cD_s}}{GL}\,.
\label{eq:condition_sur_alpha}
\end{equation}

In the case of salt, the gradient is advected along the vertical separatrix and then dissipated (see Section \ref{sec:salt}); what about colloids?
The gradient of colloids is created in the vicinity of the vertical separatrix and then advected in the cell over a distance $d$ for a time $\tau^c_{adv}\sim d/u_0$, until it is dissipated by diffusion, which happens roughly at time $\tau^c_\mathit{diff}\sim \ell_c^2/D_c$.
We obtain the travelled distance $d$ before dissipation by equaling the two times, \textit{i.e.} 
\begin{equation}
d\sim\frac{D_cD_s}{\alpha^2 G^2L} \,,
\label{eq:condition_gradient_C_diffusio}
\end{equation}
where we have used equation \eqref{eq:ell_c} for $\ell_c$.
Owing to condition (\ref{eq:condition_sur_alpha}), this distance verifies
\begin{equation}
    d\lesssim L\,. 
\end{equation}
In practice, because the gradient is present along the entire vertical, it is advected along the upper edge of the cell, then rapidly diffuses. 
If the condition \eqref{eq:condition_sur_alpha} is not met, and if the diffusiophoretic effect is weak compared with contraction by the flow field, then the phenomenology for colloids goes back to that of salt where the equilibrium state corresponds to a balance between advection and diffusion with $(a)\sim(b)$, and the inhomogeneity is located only on the vertical separatrix. 
Therefore, the colloidal heterogeneity would essentially be dissipated somewhere on the upper edge or rapidly after, depending on how the left- and right-hand side of condition \eqref{eq:condition_sur_alpha} compare. 
This is indeed what we observe in the Eulerian simulations when varying the different parameters $\alpha$, $\Pen_s$ and $\Pen_c$.  
\begin{figure}
    \centering
    \includegraphics{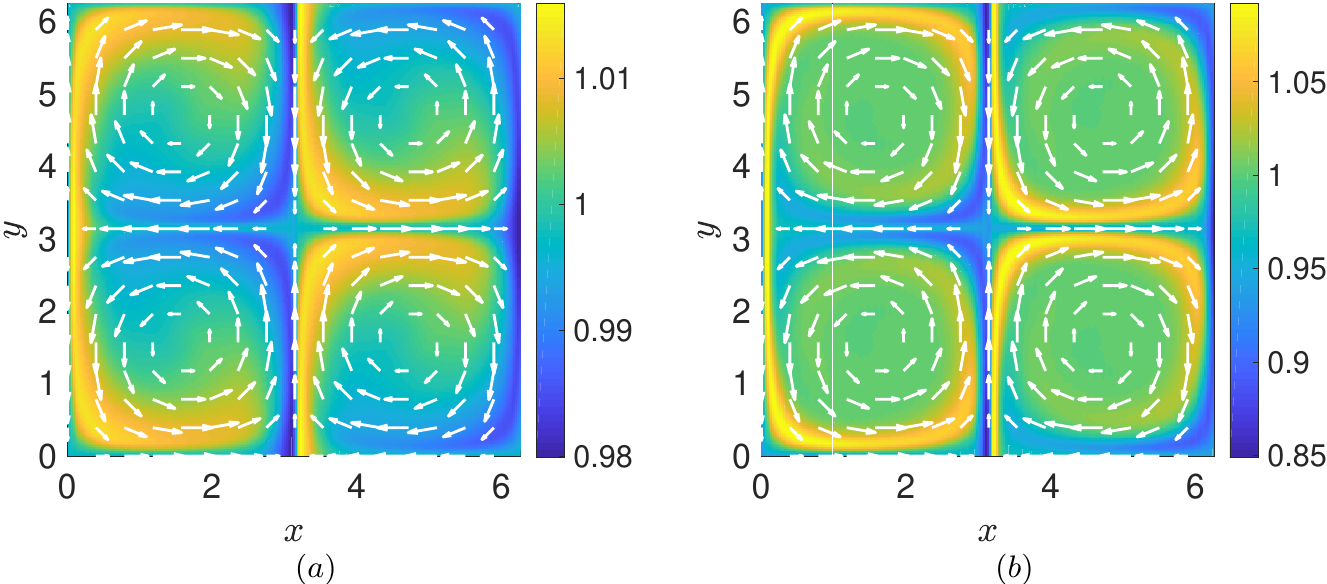}
    \caption{Eulerian colloid field $c(y,y)$ in four adjacent cells for $\alpha=10^{-3}$ and $\Pen_c=3\,141$: $(a)$ $\Pen_s=31.4$, advection and diffusiophoresis have comparable strength in the creation of gradients, the two terms in relation \eqref{eq:condition_sur_alpha} are of the same order of magnitude; $(b)$ $\Pen_s=314$, diffusiophoresis is more efficient and relation \eqref{eq:condition_sur_alpha} is met.
    \label{fig:col_eul_concentration}}
\end{figure}
As an illustration, Figure \ref{fig:col_eul_concentration} shows two typical colloidal concentration fields for $\alpha=10^{-3}$ and $\Pen_c=3\,141$ and two different salt  P\'eclet numbers $\Pen_s=31.4$ $(a)$ and $\Pen_s=314$ $(b)$: 
in figure \ref{fig:col_eul_concentration}\,$a$, the two terms in equation \eqref{eq:condition_sur_alpha} are of the same order of magnitude, whereas condition \eqref{eq:condition_sur_alpha} is satisfied in figure \ref{fig:col_eul_concentration}\,$b$. 
Note the different scales in the colourbars: except in the cores of cells which seem more homogeneous, the heterogeneities are more pronounced when the diffusiophoretic effects are stronger, with more asymmetric positive and negative deviations from the mean $C_0=1$. 

\subsection{Lagrangian results: velocity and effective diffusion}
The Eulerian simulations showed how the colloids are demixed due to diffusiophoresis in the stationary regime. As a result, it is expected that their averaged velocity is modified.
In order to explore this phenomenon when $\Pen_c$ is large, we now perform Lagrangian simulations and investigate the statistics of colloids displacement at long times, denoted $\Delta\mathbf{X}_c=\mathbf{X}_c(t)-\mathbf{X}_c(0)$. 
We denote an ensemble average over trajectories by $\langle \cdot \rangle_\mathcal{L}$; 
this allows to define the mean Lagrangian velocity $\langle \mathbf{V}\rangle_\mathcal{L}$ through the relation $\langle \Delta\mathbf{X}_c \rangle_\mathcal{L}\sim\langle \mathbf{V}\rangle_\mathcal{L} \times t$ at large times.
We also measure an effective diffusion $D_\mathrm{eff}$: 
we have $\langle (\Delta X_c)^2\rangle_\mathcal{L}-\langle \Delta X_c\rangle^2_\mathcal{L} \sim2D_\mathrm{eff}^{\,x}\times t$ and $\langle (\Delta Y_c)^2\rangle_\mathcal{L}-\langle \Delta Y_c\rangle^2_\mathcal{L} \sim2D_\mathrm{eff}^{\,y}\times t$ at large times, where we have treated separately the $x$ and $y$ directions in order to take into account a possible anisotropy. 

\begin{figure}
    \centering
    \includegraphics{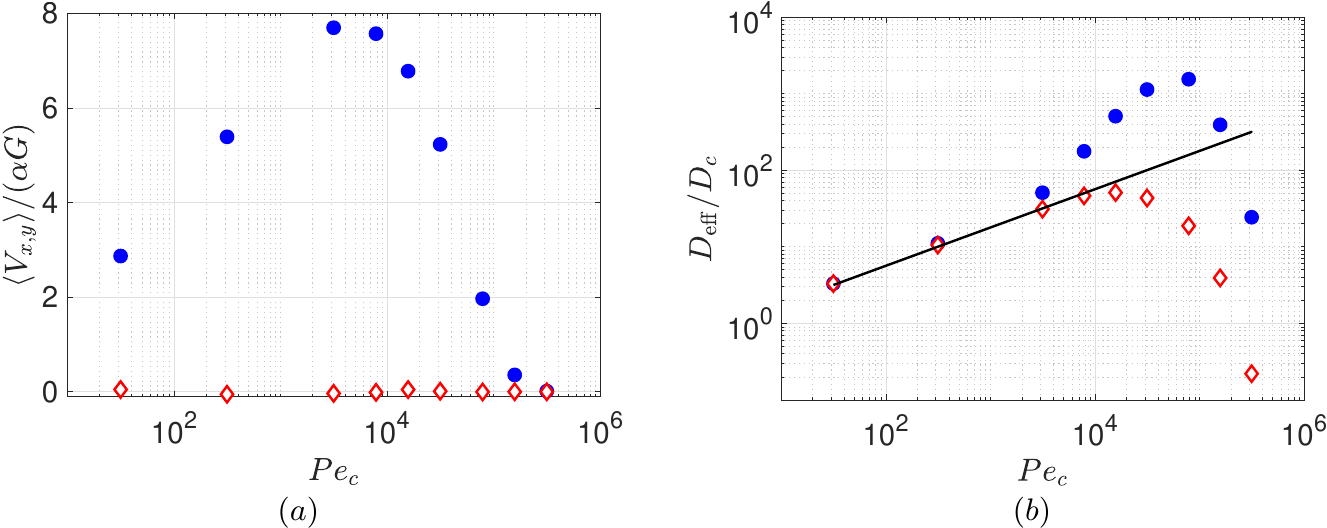}
    \caption{$(a)$ Non-dimensional mean Lagrangian velocity $\langle \mathbf{V}\rangle$ and $(b)$ non-dimensional effective diffusion $D_\mathrm{eff}$, for $\Pen_s=314$ and $\alpha=10^{-3}$ as a function of the P\'eclet number $\Pen_c$: \textcolor{blue}{$\bullet$} $x$-direction; \textcolor{red}{$\diamond$} $y$-direction. 
    The velocity is made non-dimensional using the diffusiophoretic velocity $\alpha G$ based on the mean gradient; the effective diffusion is compared with the diffusion coefficient of the colloids $D_c$. 
    The solid black line is the analytical solution proposed by \citet{bib:shraiman1987}: $D_\mathrm{eff}/D_c = Pe_c^{1/2}/\sqrt{2 \pi}$.
    \label{fig:vel_diff_lag_Pec}}
\end{figure}

Because the mean Eulerian velocity field $\langle \mathbf{u}\rangle$ is zero, one could reasonably think that the mean Lagrangian velocity $\langle\mathbf{V}\rangle_\mathcal{L}$ is of the same order of magnitude as the mean salt gradient $\alpha\mathbf{G}$. 
However, this is not at all the case when looking at figure \ref{fig:vel_diff_lag_Pec}\,$a$, that shows the ratio of mean Lagrangian $x$ and $y$ velocity components of $\langle \mathbf{V}\rangle_\mathcal{L}$ to $\alpha G$ as a function of the P\'eclet number $\Pen_c$, at fixed $\Pen_s=314$ and $\alpha=10^{-3}$: 
whereas the $y$-component remains zero as expected, the $x$-component is much larger than $1$ for most of the points. 
Moreover, the ratio has a non-trivial behaviour: at low to moderate P\'eclet numbers $\Pen_c$, the velocity increases, but eventually decreases to zero at larger P\'eclet numbers.

Figure \ref{fig:vel_diff_lag_Pec}\,$b$ shows the effective diffusion coefficient in the $x$ and $y$ directions. The line is the prediction by \citet{bib:shraiman1987} for a diffusing scalar \textit{without diffusiophoresis} in a cellular flow: 
he showed that the effective diffusion grows as the square root of the P\'eclet number, and becomes orders of magnitude higher than the diffusivity of the scalar. 
Very interestingly, we find that the effective diffusion at small to moderate P\'eclet number ($\Pen_c\le3000$) is isotropic, and follows the prediction by Shraiman, although the effects of diffusiophoresis are macroscopically visible on the velocity at those P\'eclet numbers. 
For $\Pen_c>3000$, while the mean Lagrangian velocity begins to decay (figure \ref{fig:vel_diff_lag_Pec}\,$a$), the effective diffusivity becomes highly anisotropic, with an increase in the $x$-direction compared with Shraiman's prediction, and a decrease in the $y$-direction, 
At even higher P\'eclet numbers, the effective diffusivity in both directions finally decays to zero, although for larger  P\'eclet numbers than for the velocity.

\begin{figure}
    \centering
    \includegraphics{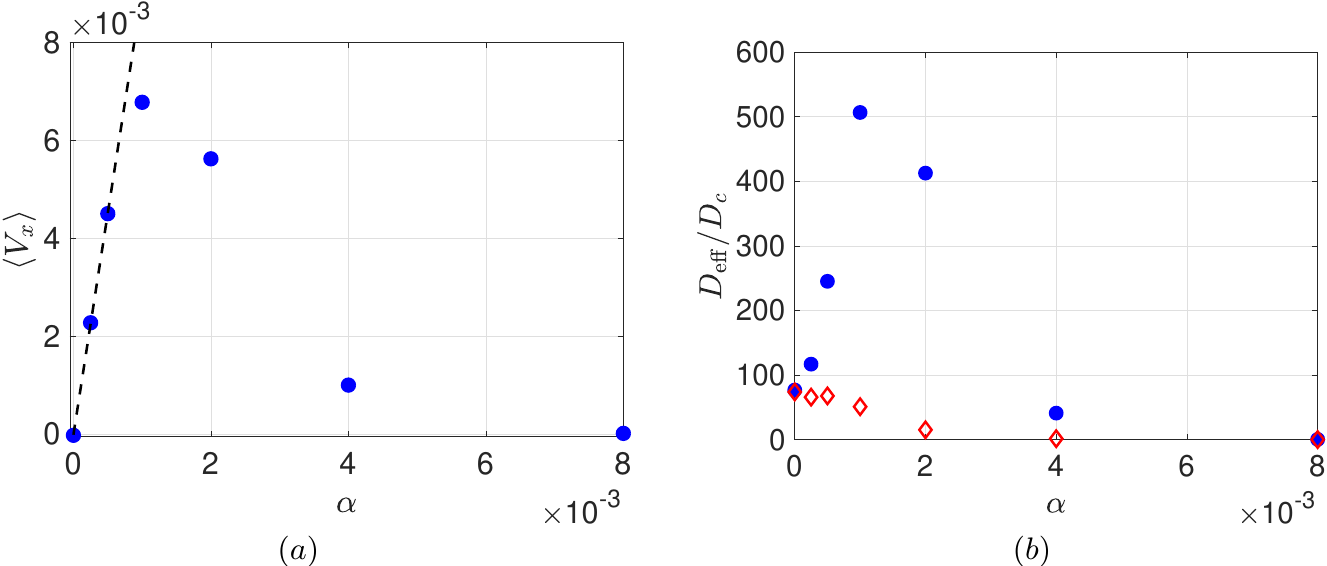}
    \caption{$(a)$ The $x-$component of the Lagrangian velocity $\langle V_x\rangle$ and $(b)$ non-dimensional effective diffusion $D_\mathrm{eff}/D_c$, for $\Pen_s=314$, $\Pen_c=15\,708$ as a function of  $\alpha$. \textcolor{blue}{$\bullet$}: $x$-direction; \textcolor{red}{$\diamond$}: $y$-direction. 
    \label{fig:vel_diff_alpha}}
\end{figure}
Figure \ref{fig:vel_diff_alpha} shows the mean Lagrangian velocity and effective diffusion as a function of the diffusiophoretic coefficient $\alpha$ for the same salt P\'eclet number and $\Pen_c=15\,708$, a P\'eclet number for which the effective diffusion is anisotropic (figure \ref{fig:vel_diff_lag_Pec}\,$b$), more realistic of the large P\'eclet numbers encountered for colloids. 
The mean velocity in the $y$-direction is again zero and is not shown here. 
For small $\alpha$, $\langle V_x\rangle$ increases linearly, before vanishing also at higher values. 
In figure \ref{fig:vel_diff_alpha}\,$b$, we recover at $\alpha=0$ the value predicted by \citet{bib:shraiman1987} for the effective diffusion in both directions. 
Then $D_\mathrm{eff}^{\,x}$ increases rapidly for small $\alpha$, before decreasing to zero at larger values; in the $y$-direction, the effective diffusivity  decreases to zero as $\alpha$ increases. 

In the next section we deepen the analysis of the colloids velocity and study the physical mechanisms at stake which explain the observations we just reported based on the Lagrangian simulations.

\section{From enhanced dispersion to blockage }
\label{sec:enhanced_disp_blockage}
\subsection{Enhanced dispersion}
\begin{figure}
\includegraphics{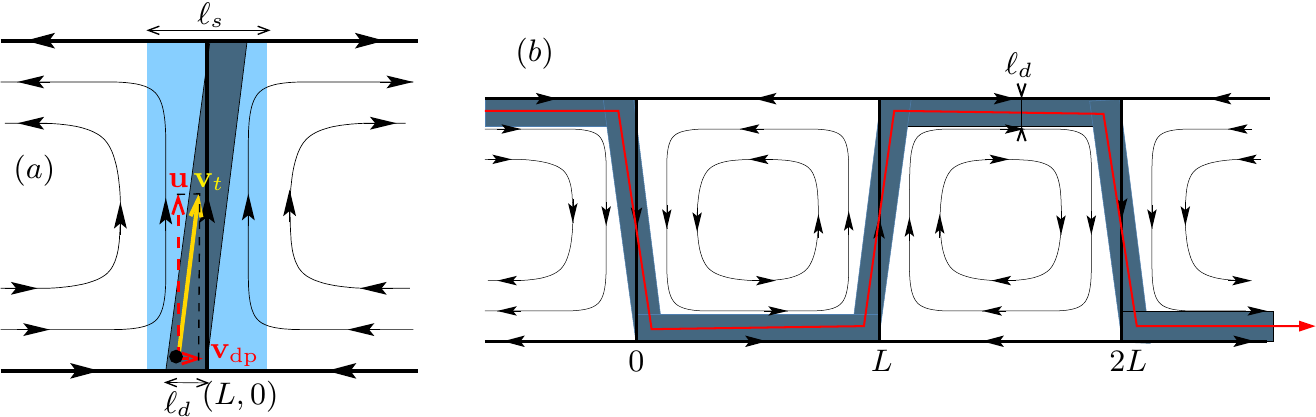}
\caption{$(a)$ Schematic displacement of a particle in the salt gradient (light blue region) around the vertical ascending separatrix $x=L$.
The Eulerian velocity field $\mathbf{u}$ is vertical ascending, and $\mathbf{v}_\mathrm{dp}$ is horizontal; 
the total displacement is therefore  $\mathbf{v}_t=\mathbf{u}+\mathbf{v}_\mathrm{dp}$, with $\mathbf{u}\approx u_0\, \mathbf{e}_y$. 
The particles that change or will change cell are located on the grey strip of width $\ell_d$. 
$(b)$ Schematic location of particles that change cell, in grey, for cells located between $x=0$ and $x=2L$; 
an example of a trajectory is shown in red. 
Outside this grey stripe, the particles remain trapped in a cell.
\label{fig:particule_traj_grad}}
\end{figure}

Figures \ref{fig:vel_diff_lag_Pec}\,$a$ and \ref{fig:vel_diff_alpha}\,$a$ showed that transport is strongly enhanced at small diffusiophoretic forcing, and blocked at higher forcing. 
In order to explain these observations, we shall return to the behaviour of colloids in the salt gradient along the vertical separatrices. 
Colloids that are in this region are shifted to the right under the effect of diffusiophoresis: 
they move with a velocity $\mathbf{v}=\mathbf{u}+ \mathbf{v}_\mathrm{dp}$ with $\mathbf{v}_\mathrm{dp}=\alpha\nabla S$. 
The velocity $\mathbf{u}$ is roughly vertical, whereas the diffusiophoretic velocity is roughly horizontal, see figure \ref{fig:particule_traj_grad}\,$a$ around the separatrix $x=L$ (upward velocity).

We denote by $\ell_d$ the horizontal displacement travelled by a colloid under the effect of the salt gradient as it rises or falls along the vertical separatrix ; 
in a first step, we suppose $\ell_d\le\ell_s$. 
Along the vertical separatrix we have $u\sim u_0 $, so that
\begin{equation}
\frac{\ell_d}{v_\mathrm{dp}}\sim \frac{L}{u_0}\,,
\label{eq:relation_ld_vdp}
\end{equation}
see figure \ref{fig:particule_traj_grad}\,$a$.
Because of equation \eqref{eq:salt_grad_separatrix} the diffusiophoretic velocity in the salt gradient can be written 
\begin{equation}
    v_\mathrm{dp}\sim \alpha G \sqrt{\Pen_s}
    \label{eq:vdp}
\end{equation}
and, finally,
\begin{equation}
\ell_d\sim\frac{\alpha G L}{u_0}\sqrt{\Pen_s}\,.
\label{eq:largeur_depletee}
\end{equation}
Hence, the particles which change vortex during their ascent (or descent) in the salt gradient are located in a strip of width $\ell_d$, indicated in grey in figure \ref{fig:particule_traj_grad}\,$a$. 
In figure \ref{fig:particule_traj_grad}\,$b$ we follow those particles on several vortices: 
in the horizontal strips, the salt gradient is negligible, so that the colloids follow the Eulerian flow field, mostly horizontal here and such that $\mathbf{u}\sim u_0\, \mathbf{e}_x$. 
Therefore, the total displacement of particles is horizontal, as found in our Eulerian and Lagrangian simulations, see for instance figure \ref{fig:vel_diff_lag_Pec}\,$a$. 
Note also that the strips coloured in grey in figure \ref{fig:particule_traj_grad}\,$b$ are in good qualitative agreement with the regions of high concentration in the Eulerian colloidal fields in figure \ref{fig:col_eul_concentration}. 
The regions not coloured in figure \ref{fig:particule_traj_grad}\,$b$ are the locations of the particles that are trapped inside a cell: 
by construction, the Lagrangian velocity of these trapped particles is zero on average. 
We can thus evaluate the spatial average horizontal velocity $V_m$ on a cell of size $L^2$:
\begin{eqnarray}
V_m&\sim&\frac{1}{L^2}\left[(\ell_d\times L)\,v_\mathrm{dp}+(\ell_d\times L)\,u_0\right]\\
&\sim&\frac{\ell_d}{L}\, (v_\mathrm{dp}+u_0)\,.
\end{eqnarray}
With equation \eqref{eq:relation_ld_vdp}, we obtain $v_\mathrm{dp}/u_0\sim \ell_d/L$, so that $v_\mathrm{dp}$ can be neglected, and, using equation (\ref{eq:largeur_depletee}),
\begin{equation}
    V_m\sim \alpha G \sqrt{\Pen_s}\,.
    \label{eq:Vm_valeur_overestimated}
\end{equation}

This simple result leads to some remarks.
\begin{itemize}
    \item[$-$] The mean colloids velocity is predicted to be much larger than that corresponding to the mean salt gradient ($\alpha G$), in accordance with figure \ref{fig:vel_diff_lag_Pec}\,$a$. 
    \item[$-$] We recover the proportionality in $\alpha$ observed in figure \ref{fig:vel_diff_alpha}\,$a$ for small $\alpha$.
    \item[$-$] Relation \eqref{eq:Vm_valeur_overestimated} is expected to slightly overestimate the actual velocity. 
    Indeed, we assumed $u\sim u_0$ on the separatrices; however, the mean of $u_0\sin kx$ on the side of a cell is rather $u_0/\pi$; changing $u_0$ into $u_0/\pi$ leads to
    \begin{equation}
    V_m\sim \alpha G \sqrt{\frac{\Pen_s}{\pi}}\,.
    \label{eq:Vm_valeur_theorique}
    \end{equation}
    Anyway, the velocity of a colloid is actually still lower when not laying exactly on the separatrix, so that even equation \eqref{eq:Vm_valeur_theorique} slightly overestimates the velocity. 
\end{itemize}

\begin{figure}
    \centering
    \includegraphics{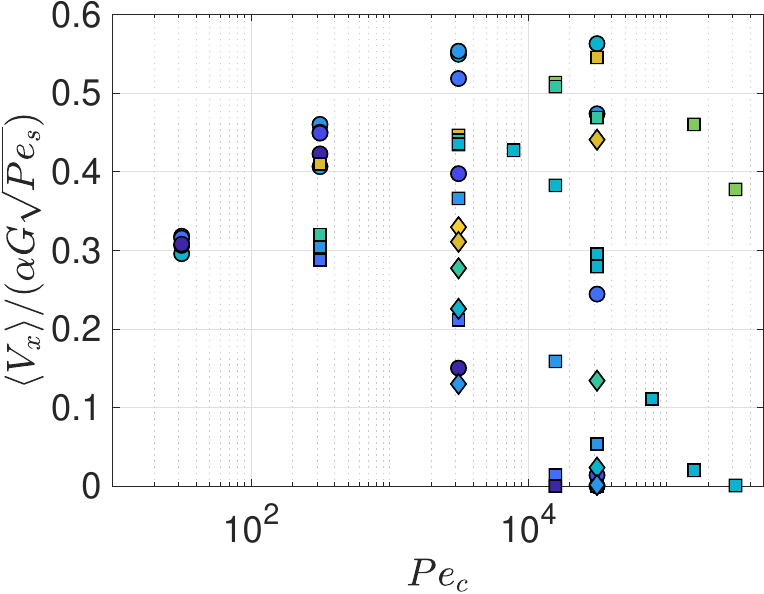}
\caption{Normalised horizontal velocity $\langle V_x \rangle_\mathcal{L}/(\alpha G \sqrt{\Pen_s})$ as a function of $\Pen_c$ for the different numerical cases performed. Symbols: $(\circ)$ $\Pen_s=31.4$; $(\square)$ $\Pen_s=314$; $(\diamond)$ $\Pen_s=3144$. The colour of symbol codes for the value of $\alpha \in [5\times 10^{-6}, ~1.6\times 10^{-3}]$, the darker the larger the value.}
    \label{fig:velocity}
\end{figure}

Let us test this prediction by performing Lagrangian numerical experiments for many different values of $D_s$, $D_c$ and $\alpha$. 
For all those values we measure the $x$-components of the velocity and display $\langle V_x \rangle_\mathcal{L}/(\alpha G \sqrt{\Pen_s})$ as a function of $Pe_c$ in figure \ref{fig:velocity}. 
It can be observed that the formula predicts the correct order of magnitude for the maximal value of the velocity ($1/\sqrt{\pi} \simeq 0.56$) at a given colloid P\'eclet number. 
However, the expression \eqref{eq:Vm_valeur_theorique} does not depend on $\Pen_c$ disagreeing with figure \ref{fig:vel_diff_lag_Pec}\,$a$, and fails to predict the rapid decay of the mean velocity observed in figures \ref{fig:vel_diff_lag_Pec}\,$a$ and \ref{fig:vel_diff_alpha}\,$a$ when increasing $\Pen_c$ and $\alpha$. 
There is thus a need to refine the analysis and figure out why the dispersion of the particles stops, both for the velocity and the effective diffusion.




\subsection{Blockage}
\label{sec:blockage}

As described previously, the grey stripe in figure \ref{fig:particule_traj_grad}\,$b$ is the location of particles that can change cell, whilst particles outside of this stripe remain trapped in a cell. 
Therefore, blockage occurs when the grey stripe becomes empty of colloids. 
In order to understand how this can happen, let us first come back to the salt profile already described in section \ref{sec:salt}. 
The horizontal salt gradient is located in the vicinity of the vertical separatrices; 
because the mean gradient $\mathbf{G}$ points toward positive $x$, one goes from low salt concentrations to high salt concentrations when crossing vertical separatrices from left to right, see figure \ref{fig:salt_eulerian}. 
As explained previously, the horizontal salt gradient at $x=0$ is built by bringing (from right to left) the scalar to the hyperbolic point $(0,L)$, on the high salt concentration side, by contraction by the velocity field along the horizontal separatrix: this implies that this separatrix should correspond to a local maximum of salt concentration. 
Symmetrically, the horizontal separatrix that relates point $(0,0)$ to $(L,0)$ builds the salt gradient at $x=L$, on the low concentration side, by bringing a deficit of salt, and should correspond to a local minimum. 
\begin{figure}
\includegraphics{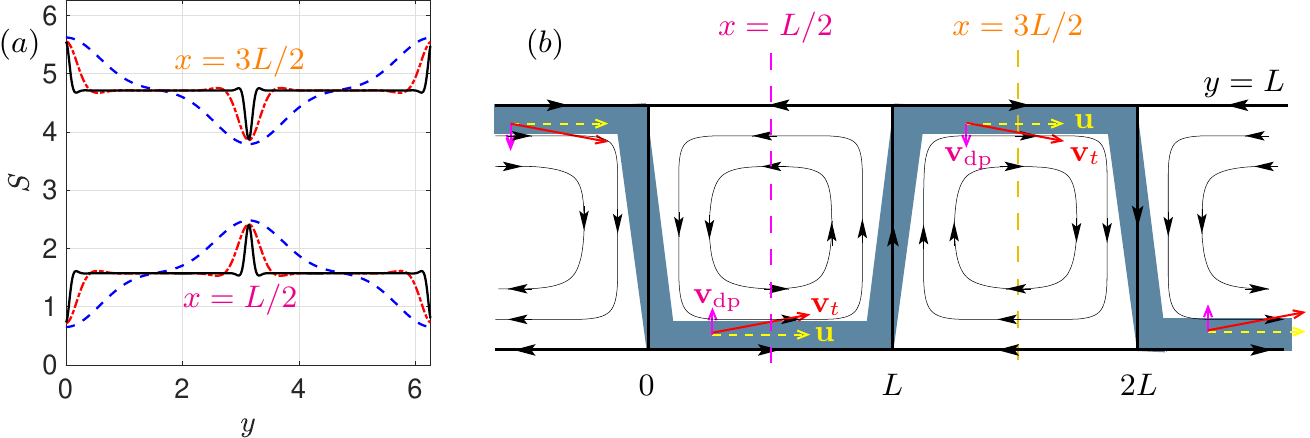}
\caption{$(a)$ One-dimensional cut of the salt concentration field along the vertical lines $x=L/2$ and $x=3L/2$ with $L=\pi$ (dotted lines in figure  \ref{fig:particule_trap}\,$b$) in the whole computational domain $[0,2L]^2$ for three different salt P\'eclet numbers: -~-~-, $\Pen_s=31.4$; -$\cdot$-$\cdot$-, $\Pen_s=314$; \textbf{---}, $\Pen_s=3\,141$. $(b)$ The horizontal parts of the stripe correspond to regions where the salt concentration gradient points vertically inwards (figure \ref{fig:particule_trap}\,$a$). 
Although the salt gradients involved are less important than around the vertical separatrices, colloids in these places are deviated outside the grey stripe. 
The red arrows indicate schematically their total velocities $\mathbf{v}_t=\mathbf{u}+\mathbf{v}_\mathrm{dp}$. 
\label{fig:particule_trap}}
\end{figure}
 This is exactly what is observed when plotting a one-dimensional cut of the salt concentration field at $x=L/2$, as shown in figure \ref{fig:particule_trap}\,$a$ for different P\'eclet numbers (see also the whole salt concentration field for $\Pen_s=314$ in figure \ref{fig:salt_eulerian}\,$a$). 
 The situation is logically reversed when considering the one-dimensional cut at $x=3L/2$. 
 Now let us come back to the displacement of colloids, also depicted in figure \ref{fig:particule_trap}\,$b$: in the horizontal grey stripes, the salt gradient is vertical and always points inwards, and so does the diffusiophoretic velocity $\mathbf{v}_\mathrm{dp}=\alpha\nabla S$; the total velocity of the colloids,  $\mathbf{v}_t=\mathbf{u}+\mathbf{v}_\mathrm{dp}$ with $\mathbf{u}\approx u_0\,\mathbf{e}_x$, deviates the colloids inwards the vortices. 
 Therefore, when diffusiophoretic effects are strong enough, the colloids rapidly leave the grey stripe and get trapped into the vortices.  
 Note, however, that the vertical salt gradients involved here are smaller than the horizontal salt gradients considered previously around the vertical separatrices; therefore, when the diffusiophoretic coefficient $\alpha$ is small, the total velocity is nearly horizontal. 
 
 The process described previously explains how colloid particles can get trapped and lead to blockage. 
 With this single mechanism at stake, however, all situations considered should lead to blockage; indeed, even though the velocity is nearly horizontal in the horizontal stripes when diffusiophoretic effects are weak, the small vertical deviation should lead to blockage on very long times. 
 This is obviously not what we observe in the numerical simulations, where we also observe enhanced transport at large times. The reason is that we have up to now neglected molecular diffusion in our schematic description:
 to effectively observe a depleted zone, and eventually for it to become a real barrier to transport from one cell to another, this band without colloids must be advected all around the cell, without having the time to diffuse. 
Blockage therefore only occurs if the diffusion time of the depleted stripe, $\ell_d^2/D_c$, is large compared with the advection time over the entire perimeter of cell $4L/u_0$. 
This competition is governed by the ratio $R^2$ of the two time scales, such that
\begin{equation}
    R =\sqrt{\frac{\ell_d^2/D_c}{4L/u_0}}\,.
    \label{eq:rapport_temps_blocage}
\end{equation}
Using relation \eqref{eq:largeur_depletee} for $\ell_d$, one obtains the condition of blockage:
\begin{equation}
R =\frac{\alpha\, GL}{2\sqrt{D_c D_s}}\gg1\,, 
\label{condition_region_depletee}
\end{equation}
where we recognise $\Delta S=GL$, the jump in concentration from one cell to the other (section \ref{sec:salt}).
Note that we recover a condition similar to what we had found before (equation \eqref{eq:condition_sur_alpha}).
Interestingly, this condition is independent of the Eulerian velocity field $\mathbf{u}$, although the blockage would not occur without the velocity field.\\ 

Throughout this section, we have assumed so far that $\ell_d\le\ell_s$. 
However, using equations \eqref{eq:ell_s} and (\ref{eq:largeur_depletee}), we obtain
\begin{equation}
    \frac{\ell_d}{\ell_s}\sim \frac{\alpha G}{u_0}\,\Pen_s= \frac{\alpha GL}{D_s}\,;
\end{equation}
therefore $\ell_d$, as calculated using \eqref{eq:largeur_depletee}, may be larger than $\ell_s$. 
Because $\ell_d$ measures the deviation of colloids by diffusiophoresis, and because the diffusiophoretic velocity is zero outside the salt gradient, this implies that at most
\begin{equation}
    \ell_d\sim\ell_s
\end{equation}
(see also figure \ref{fig:particule_traj_grad}\,$a$).
We can therefore wonder whether the expressions obtained previously for the velocity and the blockage condition remain valid in this case. 
Indeed, if equation \eqref{eq:largeur_depletee} overestimates $\ell_d$, then condition \eqref{condition_region_depletee} could falsely lead to a condition of blockage. 
We show thereafter that in that case condition \eqref{condition_region_depletee} remains a valid criterion for blockage. 

As explained previously a \textit{salt} heterogeneity of width $\ell_s$ diffuses after having travelled a distance $L$ along the vertical separatrix;  
hence, a \textit{colloidal} heterogeneity of width $\ell_s$ should travel much longer in space and time before diffusing, because $D_c\ll D_s$. 
This can be checked using the same criterion as before, the ratio $R'^2$ between the diffusing time and the travelling time on the perimeter of a cell, \textit{i.e.}, taking $\ell_d=\ell_s$ in equation \eqref{eq:rapport_temps_blocage}; we obtain
\begin{equation}
    R'=\sqrt{\frac{\ell_s^2/D_c}{4L/u_0}}=\frac{1}{2}\sqrt{\frac{D_s}{D_c}}\gg 1\,,
    \label{eq:def_R_prime}
\end{equation}
where $\ell_s$ was evaluated using equation \eqref{eq:ell_s}.
Therefore, this case when $\ell_d\sim \ell_s$ also corresponds to a situation of blockage, with no transport from one cell to the other. 
Let us show that in this situation of blockage we also have $R\gg 1$:
we denote by $\ell_d^\mathit{est}$ the overestimated width of the depleted band $\ell_d$ given by relation \eqref{eq:largeur_depletee}. 
Then $\ell_d^\mathit{est}\ge\ell_d\sim\ell_s$, and according to equations \eqref{eq:rapport_temps_blocage} and \eqref{eq:def_R_prime}, $R\ge R'$. 
Thus, we have both $R\ge R'$ and $R'\gg1$, and, finally, $R\gg 1$. The criterion \eqref{condition_region_depletee} on $R$: 
\begin{itemize}
    \item[$(i)$] $R\ll1$, enhanced dispersion;
    \item[$(ii)$] $R\gg1$, blockage;
\end{itemize}
is valid for all situations.

\begin{figure}
    \centering
\includegraphics{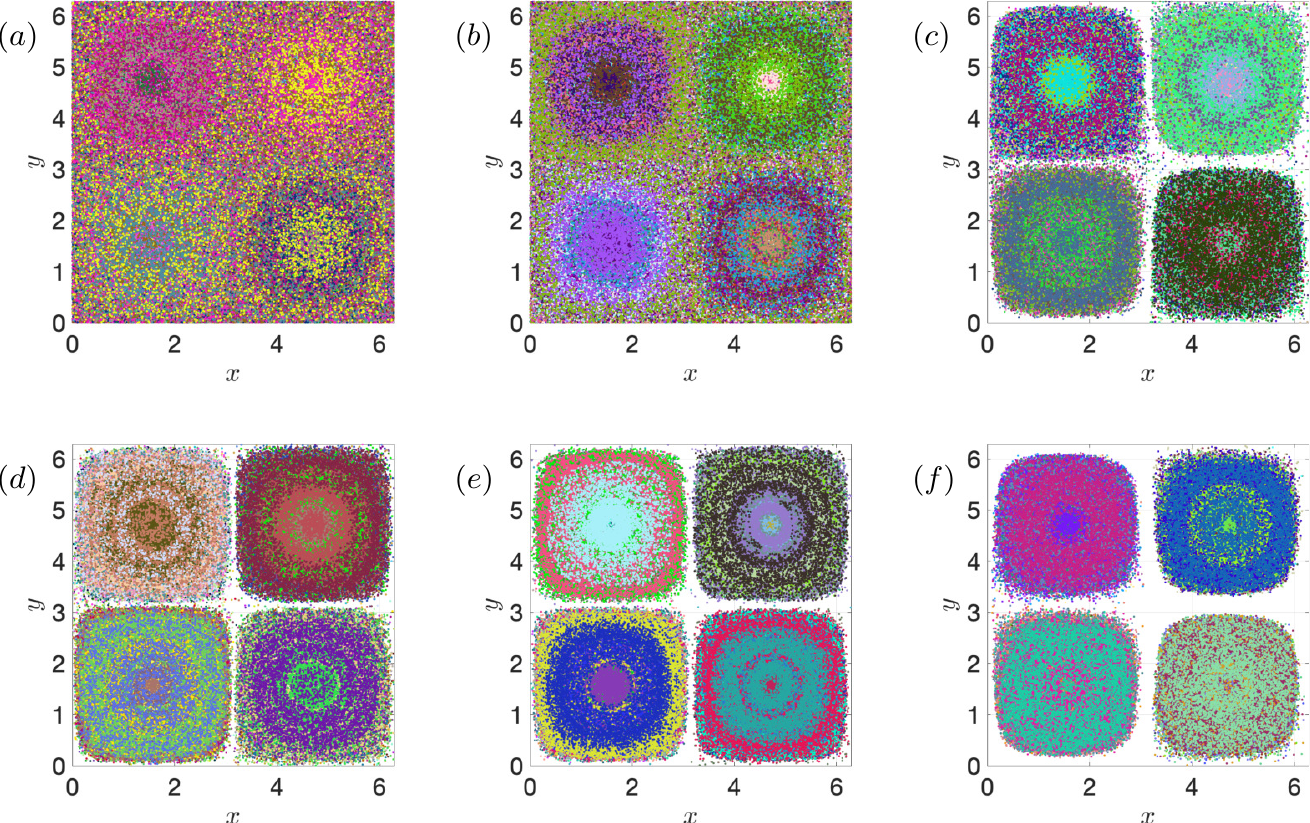}
    \caption{Successive positions of $N=100$ long trajectories of colloids, plotted modulo $2L$, in 4 adjacent cells. Particles were initially distributed uniformly and randomly inside the cells with positions dumped long after release, each colour coding for one trajectory. Parameters are $\Pen_s=314$ for each figure and different values of $\Pen_c$ and $\alpha$.
    The criterion $R$ is increasing from $(a)$ to $(f)$: 
    $(a)$ $\alpha=10^{-3}$, $\Pen_c=3\,141$, $R=0.497$; 
    $(b)$ $\alpha=10^{-3}$, $\Pen_c=15\,700$, $R=1.11$; 
    $(c)$ $\alpha=10^{-3}$, $\Pen_c= 157\,000$, $R=3.51$;
    $(d)$ $\alpha=4.\,10^{-3}$, $\Pen_c= 15\,700$, $R=4.44$;
    $(e)$ $\alpha=10^{-3}$, $\Pen_c=314\,000$, $R=4.97$; $(f)$ $\alpha=8\times10^{-3}$, $\Pen_c=15\,708$, $R=8.88$.
    \label{fig:col_lag_4cases}}
\end{figure}

We first test the blockage criterion for different values of the parameters $D_s$, $D_c$ and $\alpha$ performing simulations in the Lagrangian framework:
as a proxy for the stationary distribution of colloids, we display in figure \ref{fig:col_lag_4cases} successive positions of $N=100$ long trajectories of colloids, initially randomly and uniformly distributed inside four adjacent cells, for growing values of the criterion $R$. 
These trajectories, plotted modulo $2L$ and dumped long after release so that their position distribution is stationary,  allow to see if particles sample uniformly the cells or regroup in some localised regions. 
Although the cells are hardly distinguishable in figure \ref{fig:col_lag_4cases}\,$a$, they become more distinct as $R$ increases, and finally become completely disjoint. 
In this last regime, unfolded trajectories are trapped in cells and their mean velocity drop to zero.


\begin{figure}
    \centering
        \includegraphics{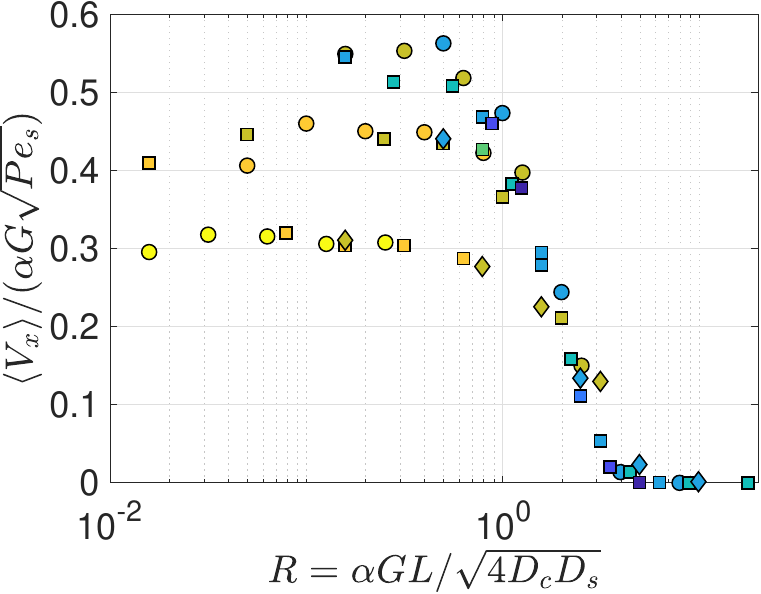}
\caption{Normalised horizontal velocity $\langle V_x \rangle_\mathcal{L}/(\alpha G \sqrt{\Pen_s})$ versus the blockage criterion $R=\alpha GL/2\sqrt{D_cD_s}$ for the same numerical cases as in figure \ref{fig:velocity}.
Symbols: $\circ$, $\Pen_s=31$; $\square$, $Pe_s=314$; $\diamond$, $\Pen_s=3144$. 
The colour of symbol codes for the value of $\Pen_c \in [31\,; 3.14 \times 10^5]$, the darker the larger the value.}
    \label{fig:velocity_rescaled}
\end{figure}

This proves that $R$ is indeed a good indicator of trapping so that we choose to plot the rescaled mean velocity $\langle V_x \rangle_\mathcal{L}/(\alpha G \sqrt{\Pen_s})$ of figure \ref{fig:velocity} as a function of the blockage criterion $R=\alpha G L/2\sqrt{D_cD_s}$ instead of $\Pen_c$. 
The result is displayed in figure \ref{fig:velocity_rescaled}: points are no longer scattered but collapse reasonably well, and fall to zero provided $R$ is large enough,  when blockage dominates over other mechanisms of transport. 
Finally, we note that the separation at small $R$ coincides with the (weak) logarithmic dependence on the colloidal P\'eclet number, as pointed out from figure \ref{fig:vel_diff_lag_Pec}\,$a$, and which was not taken into account in the present model. 
Despite this observation, the order of magnitude obtained for $\langle V_x \rangle_\mathcal{L}$ is rather good.


\section{Summary and conclusion}


We have studied the joint mixing/demixing of salt and colloids in a cellular flow with closed streamlines, and investigated how the long-term dispersion of the colloids is modified by a linear phoretic drift $\mathbf{v}_\mathrm{dp} = \alpha \nabla S$. 
In the chosen configuration of an imposed salt mean gradient along the $x$-direction $\langle \nabla S \rangle = G \,\mathbf{e}_x$, it is known that the salt concentration reaches a stationary state characterised by the presence of strong gradients localised along the vertical separatrices, so that the ratio of its effective diffusivity to the molecular diffusivity grows as the square root of the salt P\'eclet number \citep{bib:shraiman1987}.

By means of high-resolution numerical simulations performed both in the Eulerian and Lagrangian frameworks, we have shown that, starting from a uniform colloid concentration field (with no imposed colloid gradient), the colloids will demix.
We have considered the long time dynamics of this demixing process, characterised by their mean velocity $\mathbf{V}_m$ and their effective diffusivity $D_\mathrm{eff}$; 
in particular, we have studied how $\mathbf{V}_m$ and $D_\mathrm{eff}$ are influenced by the phoretic drift for a wide range of parameters $(\alpha G, \Pen_s=UL/D_s, \Pen_c=UL/D_c)$ with $\alpha$ and $G>0$.

The main finding is that we observe substantially two regimes of colloids dynamics depending on the blockage criterion $R=\alpha G L/\sqrt{4 D_cD_s}$. 
When $R<1$, the colloids are free to move and a strong demixing occurs at high $\Pen_c$; the concentration becomes homogeneous in the core of the flow cells whereas it is higher along the vertical separatrices, and in regions along the horizontal separatrices  where the velocity $\mathbf{u}$ of the flow is in the same direction as $\mathbf{G}$ ($\mathbf{G}\cdot\mathbf{u}>0$).
This is well explained as a combination of the non divergence free drift velocity, advection from the mean flow $\mathbf{u}$ and molecular diffusion. 
In this regime, the effective diffusivity is very close to $D_c$ but the mean colloids velocity is strongly enhanced in the direction of the salt gradient. Using a model which takes into account the thickness of the high concentration region with an enhanced cell-to-cell transport due to the phoretic drift, we have shown that $\langle V_x \rangle_\mathcal{L} \propto \alpha G \sqrt{\Pen_s}$. 
When $R=\alpha G L/\sqrt{4 D_cD_s} > 1$ the phoretic drift is so strong that a depletion of colloids occurs along the separatrices so that colloids can no longer migrate from one cell to another and transport is suppressed. 
In this regime of blockage, all transport observables (mean velocity, effective diffusivity) go to zero when $R$ is large enough. 
We also made an interesting observation in the transition regime $R \sim 1$ (although we could not explain it), for which the effective diffusivity along the mean gradient, $D_\mathrm{eff,x}$, first strongly increases when increasing $R$ while $D_\mathrm{eff,y}$ and $\langle V_x \rangle_\mathcal{L}$ are decreasing functions of $R$. 
Note finally that the regime of blockage, starting from a uniform distribution, is typical of compressible effects.

One may wonder if such behaviours as enhanced dispersion or blockage could be observed in practice. In the case of chemotaxis, because of the very large values of the drift coefficient $\alpha$ involved \citep{bib:chu_etal2022}, a situation of blockage would be likely to happen in such flow configurations, where the separatrices would become real transport barriers \citep{bib:berman_etal2021}. 
However, when the nutriments would lack, and the gradient $G$ would decrease sufficiently, the motile organisms would become free to leave their cell and mperhaps find some food somewhere else. The case of diffusiophoresis could be studied
in a similar configuration to that of \cite{solomon1988passive}, for which a laminar flow with closed streamlines is created in an elongated volume ($H=0.75$ cm height, $L_\mathrm{t}=15\,\mathrm{cm}$ long, $e=1.5\,\mathrm{cm}$  deep) by creating a laminar and stationary Rayleigh-B\'enard flow with a small temperature difference $\Delta T \simeq 1^\circ$ C. Whether enhanced transport or blockage would be observed in such a configuration by introducing salt and colloids at the same side (salt-in configuration) or at the opposite sides (salt-out configuration) remains an open question. 
If we choose LiCl for the salt as in \citet{bib:Maugeretal2016}, with $D_s=1360\,\mu\mathrm{m}^2$/s and a typical velocity $U\sim0.5\,\mathrm{mm/s}$, we obtain a salt P\'eclet number $\Pen_s\sim 2760$ of the same order of magnitude as the highest value used in this study. 
Considering a drift velocity of the type $\mathbf{v}_\mathrm{drift} = D_\mathrm{dp} \nabla \log S$ with  $D_\mathrm{dp}=290\,\mu\mathrm{m}^2$/s for LiCl, one may obtain an estimate of the blockage criterion $R_\mathrm{exp}$ for the experiments by replacing $\alpha$ with $D_\mathrm{dp}/S$ in $R$ to obtain $R_\mathrm{exp}=D_{dp}GL/S\sqrt{4 D_c D_c}$. 
In the case of the aforementioned configuration one would have long after salt injection $G=\Delta S/L_\mathrm{t}$ and $L=H$ so that an estimate of the criterion becomes 
\begin{equation}
R_\mathrm{exp} \simeq \frac{1}{40} \frac{D_{dp}}{\sqrt{D_c D_c}} \frac{\Delta S}{S}. 
\end{equation}
Taking roughly $\Delta S/S \sim 1$, one would then have $R_\mathrm{exp} = \mathcal{O}(0.1)$ which corresponds to enhanced transport velocity and a weak modification of the effective diffusivity. One may also argue that before developing a salt gradient over the whole width, it is developed over a much limited length at shorter times which increases the value of the criterion. One may then observe blockage at short time followed by enhanced transport at long time depending on the time delay between injection of salt and colloids. We let such experimental study for future work.

\section*{Acknowledgments}
This work was supported by the French research programs ANR-16-CE30-0028, and IDEXLYON of the University of Lyon in the framework of the French program ``Programme Investissements d'Avenir" (ANR-16-IDEX-0005).\\

\textbf{Declaration of Interests.} The authors report no conflict of interest.

\appendix
\section{Symmetry $\alpha\rightarrow -\alpha$}
\label{app:alpha}
\begin{figure}
    \centering
\includegraphics{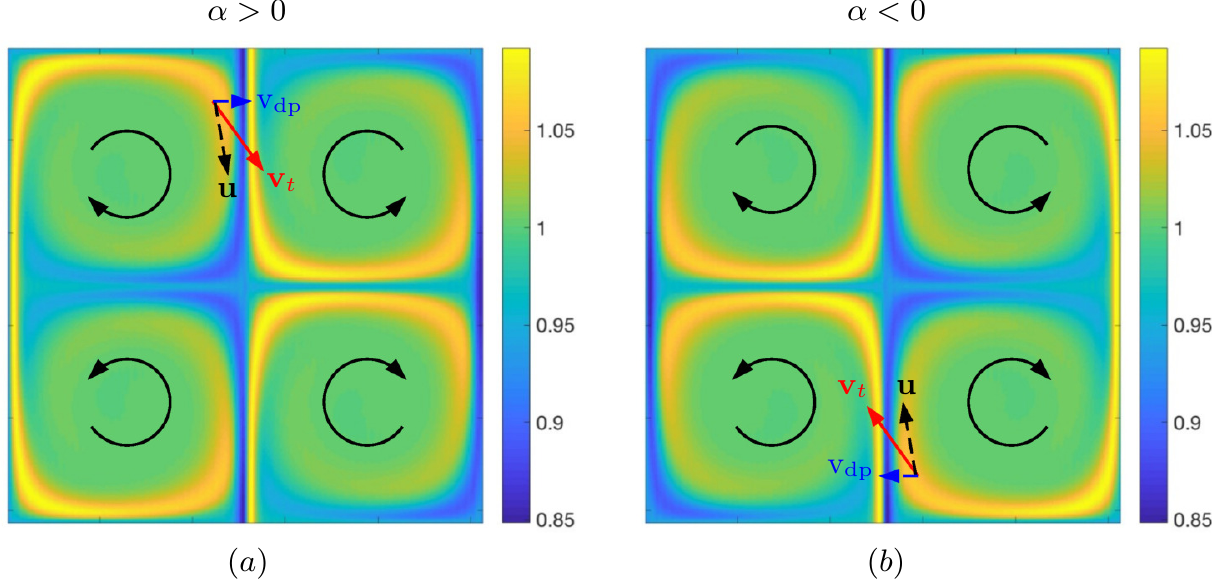}
    \caption{Colloids concentration fields on the domain $[0:2L]^2$ for $\Pen_s=314$, $\Pen_c=3141$ and opposite values of $\alpha$: $(a)$ $\alpha=10^{-3}$; $(b)$ $\alpha=-10^{-3}$. 
    Note the identical colourbar for the two cases. 
    A colloid located at a point $(x,y)$ in the case $\alpha>0$ has the opposite total velocity of the colloid located at $(mL-x,nL-y)$ where $m$ and $n$ are integers ($m=n=2$ on the figure) with $\alpha<0$. 
    }
    \label{fig:alpha_minus_alpha}
\end{figure}
Let us consider two different situations for which only the sign of $\alpha$ is changed. 
The velocity field is the same for both cases, as well as the salt concentration field. 
The equation for the  gradient of salt can be obtained from equation \ref{eq:salt}; in the stationary state considered here, it may be written:
\begin{equation}
\nabla S \cdot \mathbf{u} = D_s\nabla \cdot \nabla S \,.
\end{equation}
The velocity $\mathbf{u}$ is changed into $-\mathbf{u}$ by point symmetry around a point of coordinates $(mL,nL)$, where $m$ and $n$ are integers. 
Therefore, the gradient of salt $\nabla S$ is unchanged by this symmetry. 
Physically, a colloid which moves with a total velocity $\mathbf{v}_t=\mathbf{u}+\mathbf{v}_\mathrm{dp}$ in the case $\alpha>0$ has the opposite velocity of the symmetric colloid (with the symmetry point mentioned previously) when $\alpha<0$, as depicted in figure \ref{fig:alpha_minus_alpha}. 
This is also visible from the equation of concentration in the stationary case
\begin{equation}
    \nabla \cdot C (\mathbf{u}+ \alpha\nabla S) = D_c\, \nabla^2 C
\end{equation}
which is invariant under the point symmetries mentioned above when $\alpha$ is changed into $-\alpha$. 
As an illustration, figure \ref{fig:alpha_minus_alpha} shows two concentration fields with opposite values of $\alpha$.



\bibliographystyle{jfm}
\bibliography{Compressibility}

\end{document}